\documentclass{ar-1col-S2O}
\usepackage[numbers]{natbib}
\usepackage{url}
\usepackage{soul}
\usepackage{xspace}
\usepackage{todonotes}
\usepackage{amsfonts}
\usepackage{amssymb}
\usepackage{amsmath}
\usepackage{float}
\DeclareMathSymbol{\mh}{\mathord}{operators}{`\-}

\usepackage{tabularx,ragged2e,booktabs,caption}
\usepackage{ltablex}
\usepackage{xltabular}

\usepackage{adjustbox} 
\usepackage[version=4]{mhchem}
\usepackage{tasks}
\usepackage{float}
\usepackage{mhchem}
\usepackage{textgreek}

\newcommand{\caion}{\ce{Ca^{2+}}\xspace}

\jname{Xxxx. Xxx. Xxx. Xxx.}
\jvol{AA}
\jyear{YYYY}
\doi{10.1146/((please add article doi))}

\begin{document}

\markboth{Author et al.}{Short title}

\title{The evolution of systems biology and systems medicine: From mechanistic models to uncertainty quantification}

\author{Lingxia Qiao,$^{*}$ Ali Khalilimeybodi,$^*$ Nathaniel J. Linden-Santangeli,$^{*}$ and Padmini Rangamani
\affil{Department of Mechanical and Aerospace Engineering, University of California San Diego, La Jolla, CA; email: prangamani@ucsd.edu}
\affil{$^*$These authors contributed equally to this work.}}

\begin{abstract}
Understanding the mechanisms of interactions within cells, tissues, and organisms is crucial to driving developments across biology and medicine. 
Mathematical modeling is an essential tool for simulating biological systems and revealing biochemical regulatory mechanisms. Building on experiments, mechanistic models are widely used to describe small-scale intracellular networks and uncover biochemical mechanisms in healthy and diseased states. 
The rapid development of high-throughput sequencing techniques and computational tools has recently enabled models that span multiple scales, often integrating signaling, gene regulatory, and metabolic networks. 
These multiscale models enable comprehensive investigations of cellular networks and thus reveal previously unknown disease mechanisms and pharmacological interventions. 
Here, we review systems biology models from classical mechanistic models to larger, multiscale models that integrate multiple layers of cellular networks. 
We introduce several examples of models of hypertrophic cardiomyopathy, exercise, and cancer cell proliferation. 
Additionally, we discuss methods that increase the certainty and accuracy of model predictions. 
Integrating multiscale models has become a powerful tool for understanding disease and inspiring drug discoveries by incorporating omics data within the cell and across tissues and organisms.
\end{abstract}

\maketitle

\tableofcontents

\section{INTRODUCTION}
Over the past three decades, our understanding of cell signaling has expanded through the use of systems biology approaches, revealing the complex interactions and regulatory networks underlying cellular and tissue functions~\cite{Avi2017}. 
Integral to systems biology, mathematical and computational modeling has provided quantitative frameworks to incorporate and analyze experimental data, enabling researchers to simulate and predict regulatory mechanisms that drive dynamic behaviors of cellular systems~\cite{Ji2017}. 
The development of sequencing techniques and computational tools has facilitated the transition from systems biology to the emerging field of systems medicine by enabling more precise, personalized, and clinically relevant applications~\cite{Ho2020-ss}. 
For example, advanced bioinformatics tools have enabled the integration of omics with clinical data, providing a comprehensive view of patient responses to clinical trials~\cite{Borrego-Yaniz2024-wq}. 
However, developing predictive models for systems medicine faces multiple challenges. 
Accurately modeling drug-target interactions requires understanding the complex molecular mechanisms of drug action, but genetic, environmental, and lifestyle factors can contribute to variability and confound our understanding of drug efficacy and safety among patients~\cite{Schenone2013-cj}. 
Furthermore, modeling cross-talk between organ systems is challenging due to the need to integrate data and interactions across multiple spatial scales, from molecular and cellular levels to organ and systemic levels, and across multiple temporal scales, from milliseconds to days and weeks~\cite{Sung2019-bt,Li2015-pf}. 
These challenges can undermine the accuracy and reliability of model predictions, directly affecting clinical decision-making and the effectiveness of personalized treatments. 
Therefore, uncertainty quantification of model predictions is essential for making informed decisions in clinical research~\cite{Begoli2019}. 
Here, we review advancements in modeling approaches in systems biology and systems medicine. 
We also discuss how uncertainty quantification can improve the reliability of model predictions by reducing parametric uncertainty and aiding in model selection. 
The combination of modeling and uncertainty quantification provides opportunities to investigate large-scale cellular networks, tissues, and even organisms, shedding light on disease mechanisms and novel treatments.

\section{CLASSICAL MECHANISTIC MODELS}
To accurately capture the dynamics of cellular networks and generate reliable predictions, mechanistic models leverage experimental data that measures the biochemical activity of signaling molecules and the protein-protein interactions within cells~\cite{Review_metabolic_2014,Kitano_review_systems_biology_2002,Bruggeman2007_review_systems_biology, Kitano2002_review_systems_biology}. 
The corresponding experimental techniques that support model building include Western blotting, fluorescence microscopy, chromatin immunoprecipitation, and quantitative PCR (qPCR), among others.
These techniques provide important information on the `status' of mRNAs, proteins, and metabolites, including the concentration or level, enzymatic activity, biochemical modification state, cellular location, or binding partners.
Each of the interactions between biochemical species can be modeled as a chemical reaction.
Systems biology builds up networks of reactions and leverages dynamical systems theory to make predictions including how signaling molecules in the cell respond to extracellular stimuli, how genes are regulated by specific transcription factors, and how metabolites interact~\cite{Network_1,Network_2,Network_3}.
Many databases have been developed to collect relevant data and to provide the reaction networks for many well-studied intracellular pathways~\cite{Chowdhury2015}, including the KEGG, TRRUST, and Reactome databases.

\begin{figure}[h]
\includegraphics[width=1\textwidth]{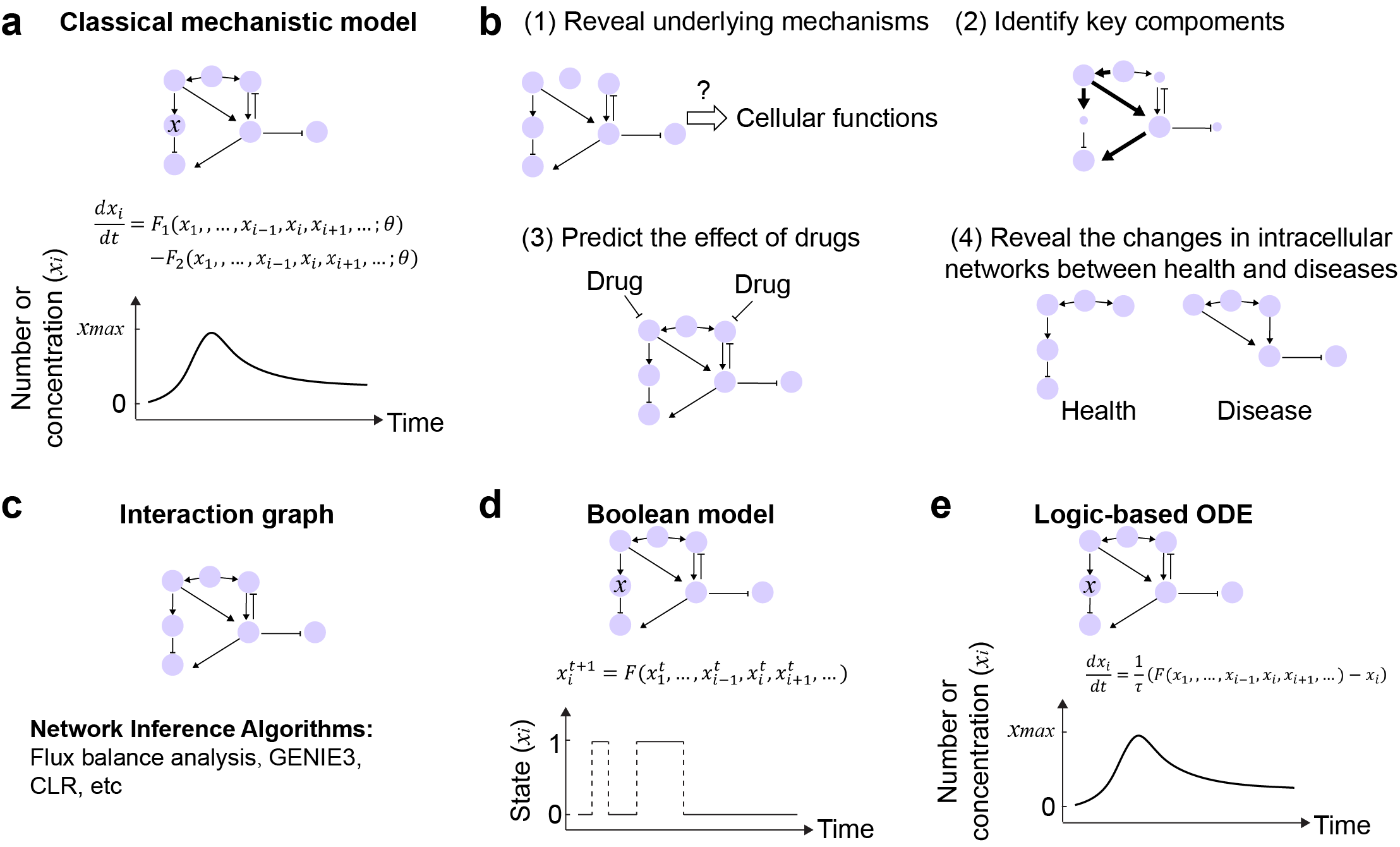}
\caption{\textbf{Classical mechanistic and network topology-based models.} (\textbf{a}) Schematic of classical mechanistic models. 
Models are ordinary differential equations that describe the dynamics of interacting biochemical molecules, where $x$ denotes the number or concentration of species. 
Furthermore, $F_1$ and $F_2$ represent the production and degradation rates, respectively, and are based on kinetics such as mass-action or Michaelis–Menten. 
(\textbf{b}) Applications of classical mechanistic models. (\textbf{c-e}) The schematic of interaction graphs, Boolean models, and logic-based ordinary differential equations (ODEs), respectively. We refer the readers to~\cite{Huynh2010_GENIE3} for GENIE3 and to~\cite{Faith_CLR} for CLR. Schematic of network is adapted from Figure 1 of Reference~\cite{Khalilimeybodi2024}.}
\label{fig:model_review}
\end{figure}

Once the network structure has been established, classical mechanistic models can be written to predict the behavior of intracellular networks. 
To capture the dynamical response for molecule $x_i$ ($i=1,2,\cdots$), ordinary differential equations (ODEs) are used and are usually written as,
\begin{equation}\frac{dx_i}{dt}=F_1(x_1,\cdots,x_{i-1},x_i,x_{i+1},\cdots;\theta)-F_2(x_1,\cdots,x_{i-1},x_i,x_{i+1},\cdots;\theta),
\label{eq:mechanistic}
\end{equation}
where $x_i$ ($i=1,2,\cdots$) denotes the number or concentrations of species (\textbf{Figure \ref{fig:model_review}A}). 
$F_1$ and $F_2$ represent the production and consumption rates, respectively.
The formulation of $F_1$ and $F_2$ are based on the kinetics of included chemical reactions and often mass action or Michaelis-Menten kinetics are used to prescribe rate laws.
Furthermore, $\theta$ represents all of the kinetic parameters, including, for example, rates of dissociation, degradation, and production.
Provided the full system of ODEs describing the evolution of all species in the system with specified initial conditions, the solution of ODEs mimics the dynamics of biochemical molecules, generating a prediction for temporal response in the intracellular network.
Importantly, ODE-based models rely on several strong assumptions about the nature of the biochemical systems, for example, that concentrations are well-mixed in the cellular environment, that molecules have negligible volume, and that biological noise does not impact the dynamics.
Thus, other types of classical mechanistic models are required when these assumptions do not hold.
For example, partial differential equations can capture the heterogeneous distributions of biochemical molecules in space~\cite{Azeloglu2015_review_partial,Kholodenko2006_space}; particle-based models enable inclusion of the structural organization of molecules~\cite{Dignon2020_LLPS}; and, stochastic ODEs are able to mimic the dynamics of species in the presence of biological noise~\cite{Tsimring_2014_noise}.
While these systems are mathematically more complex, several software packages are available for spatial, stochastic, or particle-based simulations~\cite{Schaff1997_vcell1,Laughlin2023,LAMMPS}.

Parameter estimation is essential to constrain classical mechanistic models to available data.
Historically, the values of kinetic parameters (e.g., $\theta$ in equation \ref{eq:mechanistic}) within the cell were estimated \textit{in vitro}, however, due to the highly connected intracellular networks and complicated cellular environment, these estimates proved inaccurate~\cite{Para_vitro_1,Para_vitro_2,Para_vitro_3,Para_vitro_4}.
Newer methods have improved out ability to estimate kinetic parameters \textit{in vivo}, for example diffusion rates can be estimated from fluorescence recovery after photobleaching (FRAP) assays~\cite{Day2012,Pincet2016}, protein interaction dissociation constant from F\"{o}rster resonance energy transfer (FRET) assays 
\cite{Liao2015}, fluorescence correlation spectroscopy~\cite{Sudhaharan2009_KD} or nuclear magnetic resonance (NMR) experiments~\cite{Fielding2007}, protein concentrations/or abundances from fluorescence microscopy~\cite{Wu2005,Coffman2012,Joglekar2008}, and ligand-binding kinetics from surface plasmon resonance (SPR)~\cite{Patching2014_SPR}.
However, these methods remain limited in their ability to measure every kinetic parameter within a system, leaving many unknowns in a model.
In addition to hindering predictive performance, this inconsistency also causes the unexpected \textit{in vivo} behavior of drugs even when the drugs exhibit good \textit{in vitro} performance~\cite{Para_vitro_3,Para_vitro_drug,Para_vitro_drug_1}, slowing down drug discovery.

As a complementary method, computational researchers have developed several parameter estimation algorithms, including frequentist~\cite{molesParameterEstimationBiochemical2003,Ashyraliyev2008} and Bayesian approaches~\cite{Wilkinson2007,Toni2008,Klinke2009,Golightly2011,Ghasemi2011,Liepe2014,gerisUncertaintyBiologyComputational2016,Valderrama2019,Mortlock2021,Linden2022} to infer the value or distribution of unknown parameters from available data.
In the frequentist approaches, parameters are estimated by solving an optimization problem, whose objective function typically measures the difference between model output and experimental data.
This approach is implemented in use-friendly systems biology software, including COPASI~\cite{COPASI}, Data2Dynamics~\cite{raueData2DynamicsModelingEnvironment2015}, and VCell~\cite{Schaff1997_vcell1,COWAN2012195_Vcell2}, or can be implemented manually by directing defining the objective function and optimization algorithm.
Numerous studies on dynamical systems modeling have employed frequentist approaches to fit the experiment data (see~\cite{Orton2008,Mendes1998,alon2019introduction,Schoeberl2002,Khalilimeybodi2023,Ohadi2019,Qiao2023_MSB} for more examples). 
Frequentist approaches estimate the uncertainty in parameter estimates by computing the confidence intervals around the optimal value of parameters~\cite{moon2005mathematical, Ashyraliyev2008}.
In contrast to frequentist methods, Bayesian approaches consider the unknown parameters as random variables and characterize corresponding probability distributions conditioned on available data (also called posterior distributions) by leveraging Bayes’ rule~\cite{gelman2013bayesian, smith2013uncertainty}.
In addition to the frequentist and Bayesian approaches, other approaches have also been developed to refine parameter space iteratively, such as \textit{CaliPro}~\cite{joslynCaliProCalibrationProtocol2021} or virtual population approaches~\cite{allenEfficientGenerationSelection2016, riegerImprovingGenerationSelection2018}.

Mechanistic models have been widely used in the field of systems biology to reveal the underlying mechanisms of how cells execute various biological functions (\textbf{Figure \ref{fig:model_review}B})~\cite{alon2006introduction,annurev2013,Kestler2008_review_model,Jong2002__review_model}.
Specifically, scientists have built classical mechanistic models for well-known signaling pathways, gene regulatory networks, and metabolic pathways.
Some key applications include: signaling pathways that regulate cell proliferation, apoptosis, inflammation response, metastasis, differentiation~\cite{Klipp2006_review_S} (e.g., EGFR~\cite{Schoeberl2002}, NF\textkappa B~\cite{LIPNIACKI2004_NFkb}, \caion~\cite{Dupont2011_Ca}, cAMP~\cite{Ohadi2019}, PI3K~\cite{Pappalardo2016_PI3K}, ERK~\cite{Schoeberl2002}, YAP/TAZ~\cite{Khalilimeybodi2023} pathways); gene regulatory networks that control embryo development, circadian clocks, cell differentiation, and epithelial–mesenchymal transition~\cite{Hong2015_EMT,Ukai-Tadenuma2011_Circadian,Levine_embro,Narula_diffe}; and metabolic pathways that regulate the metabolism of adenosine triphosphate (ATP), glucose, cholesterol, amino acid, and retinoic acid~\cite{Nielsen_review_meta}.
With these models, researchers have revealed the underlying mechanisms of cellular functions by answering the following questions: i) what is the core motif that drives observed behaviors? ii) what is the effect of specific reactions or molecules? iii) are the cellular functions robust to changes in kinetic parameters? 
Because of the limitations of molecular techniques, not all hypotheses can be tested to identify the right mechanisms, but this shortcoming can be overcome by using the model mainly through the \textit{in silico} perturbation of molecular activities and protein-protein interactions.

Classical mechanistic models also provide valuable insights into systems medicine because disease states are closely related to aberrant cellular functions.
Specifically, models help to identify key components that drive diseases, predict the drug effects, and reveal the changes in intracellular networks between health and disease states (\textbf{Figure \ref{fig:model_review}B})~\cite{Sun2017_mechanistic_review,Vakil2019_mechanistic_review}.
The most common approach to investigate these components is to use a model to predict the effects of perturbations to kinetics parameters or variations in the input stimulus to the model.
The impacts of these \textit{in silico} studies span many aspects of cellular biology.
For example, in understanding cellular behavior, modeling studies have revealed the effects of changes in enzyme binding kinetics on apoptosis pathway dysfunction~\cite{Classical_cancer_1}, the role of PTEN protein expression in resistance to anti-HER2 cancer therapies~\cite{Faratian2009_PTEN}, and the importance of crosstalk between signaling pathways in the relative sensitivity to drugs~\cite{Sun2016}.
Additionally, detailed mechanistic models can also lead to an improved understanding of disease mechanisms by predicting the changes in molecular activities that drive disease progression~\cite{Sun2013_stress,Sommariva2021_effects,Su2024_CFTR}.
Finally, models can predict drug effects and optimal drug combinations to guide the design of novel therapies with improved therapeutic effects~\cite{Sun2013_stress,Classical_ErbB3}.

Despite the extensive use of mechanistic models in systems biology and systems medicine, the rapidly increasing amount of experimental data on intermediate reactions, crosstalk among multiple signaling pathways, gene regulatory links, cellular localization of molecules, and cell-cell communication has brought new challenges.
Detailed models often have limited capacity to integrate diverse and multiscale datasets, such as genomics, proteomics, and clinical data, partly due to the need for significant computational power and advanced algorithms to handle such data~\cite{Stadter2021-ab,Alyass2015-nq}. 
Furthermore, new challenges are encountered as model sizes grow to consider more experimentally validated biochemical reactions. 
Estimating parameters for large-scale models dramatically increases the computational cost of both parameter estimation and during the selection of adjustable search parameters~\cite{Penas2017-pj,Gabor2015-gq}.
Furthermore, since different cellular processes may span multiple time scales---for example, signal transduction occurs in seconds to minutes while gene regulation happens over hours---modeling the coupling between slow and fast reactions by ordinary differential equations (ODEs) leads to stiff systems that can pose new numerical difficulties to solve~\cite{Fletcher2022-ro}.
 
Taken together, classical mechanistic models faithfully capture the kinetics of biochemical reactions and generate quantitative predictions that can be constrained to experiments.
Thus, classical systems biology models provide reliable predictions of the mechanisms of normal cellular functions, drug efficacy, and disease. 
Nevertheless, these models have usually been developed for small-scale intracellular networks and are limited in their ability to make multi-scale predictions due to the above challenges in handling large amounts of experimental data.

\section{NETWORK TOPOLOGY-BASED MODELS}

Instead of building classical mechanistic models by experimentally identifying each reaction and then estimating kinetic parameters, network topology-based approaches, offer an alternative modeling framework that does not require precise fitting of kinetic parameters.
The three typical approaches in this category are interaction graphs, Boolean networks, and logic-based ODEs. 
Here, we briefly review these approaches and discuss how they can be applied to systems medicine.

Interaction graphs are composed of nodes and edges, which represent biochemical molecules and regulatory links, respectively (\textbf{Figure \ref{fig:model_review}C}).
With the advance of omics data, bioinformatic tools have been extensively developed to infer the network of interactions in a system at a large scale, especially for metabolic and gene regulatory networks. 
The inference of metabolic networks started with only metabolic reactions and then expanded to the genome-scale by adding gene-protein-reaction rules based on the annotations of all genes.
As the name implies, genome-scale metabolic networks contain reactions among metabolites and related enzymes whose activities or levels are regulated by gene expression.
Currently, genome-scale metabolic networks have been constructed for a variety of organisms~\cite{Palsson2007_meta,Watson_review_meta,Nielsen_review_meta}.
Furthermore, genome-scale metabolic networks are further reconstructed to make them applicable for cells in different tissues or under disease states within the same organism~\cite{Robaina2014_metabolic_review,Foguet2022_organ_meta}, since not all metabolic reactions take place when the cell context changes.
The major algorithms for such reconstructions are based on flux balance analysis, which calculates the steady-state flow of metabolites by optimizing the phenotype (e.g., biomass production) under the quasi-steady-state assumption~\cite{Palsson2010_FBA}.
Therefore, genome-scale metabolic networks do not require kinetic parameters for each metabolic reaction.
Though genome-scale metabolic networks do not predict the dynamic behavior of metabolites and enzymes, they are extensively used in the field of systems medicine~\cite{Lewis2012_review_meta}, for example, to understand metabolic strategies under different nutrition conditions~\cite{Elsemman2022_meta}, to identify functional metabolic shifts in disease states~\cite{Dougherty2021_shift,Galhardo2013}, and to predict biomarkers of diseases~\cite{Blais2017_biomarker,Galhardo2013}.
These applications often require high-throughput data such as genomics and proteomics data.

Gene regulatory networks, often inferred directly from gene expression profiles, have also emerged as a useful tool in systems biology and systems medicine.
These inferred networks help to identify the interaction among key genes in a specific context because it is not feasible to experimentally study each pair of gene interactions given the large number (nearly 19,000 \cite{Ezkurdia2014_gene_num}) of genes in the human genome.
Similar to interaction graphs, gene regulatory networks represent each gene by a node in the graph and each regulatory interaction by an edge.
Depending on the chosen network inference algorithm, edges can be directed or undirected to reflect causality between interactions, signed or unsigned to suggest the direction of interactions, and weighted or unweighted. 
Algorithms to infer gene regulatory networks can be categorized based on the underlying methodology and included approaches that leverage estimation of correlations, regression analysis, Bayesian inference, and information theory~\cite{Saint2020_GRN_review,Huynh-Thu20190_GRN_review}. 
Network inference has broad application in systems medicine~\cite{Emmert2014_GRN_review}. 
First, it enables the discovery of key regulators for cellular functions~\cite{Kamimoto2023_grn_regulator}.
Second, it can identify biomarkers that are related to diseases~\cite{Gladitz2018_gene_candidate,Mohammadi2018_biomarker}. 
Third, it helps predict the genes that should be targeted to effectively treat diseases with a genetic basis~\cite{Madhamshettiwar2012_drug_target,Vundavilli2019_drug_target}.

Compared with the metabolic and gene regulatory networks, signaling networks are harder to infer.
One reason is that the protein-protein interactions depend strongly on post-translational modifications of proteins, which can exhibit variability across distinct cell types and developmental stages~\cite{Beltrao2013_PTM,Lee2023_PTM}.
Therefore, the construction of signaling networks is usually based on experimental data and literature search. 
Similar to the construction of metabolic and gene regulatory networks from omics data mentioned above, signaling networks inferred using interaction only provide a static description of the connectivity between involved species. 
Although the resulting models are unable to predict the dynamic behavior of signaling networks, they help to elucidate the mechanisms underlying cellular functions by using network analysis tools such as clustering, link prediction, perturbation, and network alignment~\cite{Samaga2013_review_model,Ma2009_graph_review,Koutrouli2020_graph_review,Pavlopoulos2011_graph_review}. 
One example of how an inferred signaling network can shed light into a biological system comes to from the work of Ma'ayan et al.~\cite{Avi2005_plasticity}, who inferred a signaling network to better understand neuronal homeostasis and plasticity.
By identifying the highly connected proteins and then calculating the number of involved regulatory motifs, the authors suggested that the highly connected proteins, including mitogen-activated protein kinase (MAPK), calcium-calmodulin-dependent protein kinase II (CaMKII), protein kinase A (PKA), and protein kinase C (PKC), play important roles in determining the neuron's choice between homeostasis and plasticity.  

While interaction graphs provide a static understanding of a signaling network, Boolean models and logic-based ODEs go beyond inferring the connectivity between species and approximate the systems' dynamics behavior.
In a Boolean model, the biochemical species are assumed to have only two states: zero denoting an absent or inactive state, and one denoting a present or active.
The general form of Boolean models is written as
\[x_i^{t+1}=F(x_1^t,\cdots,x_{i-1}^t,x_i^t,x_{i+1}^t,\cdots),\]
where $x_i^t$ ($i=1,2,\cdots$) denotes the state of biochemical molecules at time $t$, and can only be 0 or 1 (\textbf{Figure \ref{fig:model_review}d}). 
Here, $F$ describes how other biochemical molecules change the state of $x$ and is often a phenomenological function.
In contrast to Boolean models, logic-based ODEs are continuous in both the state of biochemical molecules and in time. 
The general form of logic-based ODEs is as follows
\[\frac{dx_i}{dt}=\frac{1}{\tau}\Big(F(x_1,\cdots,x_{i-1},x_i,x_{i+1},\cdots)-x_i\Big),\]
where $x_i$ ($i=1,2,\cdots$) denotes the number or concentration of the biochemical molecules (\textbf{Figure \ref{fig:model_review}e}).
In the model, $\tau$ is the time scale, and $F$ denotes the production rates caused by other  biochemical molecules.
The form of $F$ can be piecewise-linear, polynomial, sigmoid, or Hill functions~\cite{Wittmann2009_transform,Krumsiek2010_transform_software,Mendoza2006_discrete_conti,Samaga2013_review_model}.
Oftentimes, Hill functions are normalized to ensure the range of $F$ is between 0 and 1~\cite{Kraeutler2010}.
Normalized Hill equations can improve quantitative predictions of functional relationships within the network compared with other logic-based approaches~\cite{Kraeutler2010}. 
Logic-based ODEs provide a phenomenological model distinct from a classical mechanistic representation because $F$ is assumed to be a phenomenological functional form; however, similar to mechanistic ODEs, logic-based ODEs can also predict dynamic behavior and thus require the estimation of kinetic parameters.
Thus, the tradeoff between the large amount of experimental data and small-scale networks for classical mechanistic models can be partially mitigated by using logic-based ODEs, because these phenomenological models can predict the dynamic behavior without the details of binding partners and interaction kinetics. 

Boolean models and logic-based ODEs can also be used to predict the effect of new drugs and to identify key components in disease~\cite{Traynard2017,Hemedan2022_review_Boolean,Bloomingdale2018_review_Boolean,Zhang2014_logic_review,Morris2010_review_logic}.
Predicting drug effects is usually obtained by perturbing the corresponding drug targets and then simulating the perturbed system. 
Due to the relative simplicity of logic-based models, these two approaches are applicable for modeling large-scale biological systems, such as multiple signaling pathways with crosstalk, gene regulatory networks with a huge number of genes,  metabolic networks that execute several functions, or a combination of the above three types of networks.
Therefore, predicting drug effects based on these two models allows consideration of not only the crosstalk among intracellular networks but also numerous subsequent processes.
Owing to this advantage, these two approaches have been widely used to predict drug effects and optimal drug combinations in many diseases, for example, hypertrophic cardiomyopathy~\cite{Tan2017}, rheumatoid arthritis~\cite{RA_2021}, and several cancers~\cite{Cancer_1,Cancer_3,Zhang2008_leukemia_target}.
In addition to the prediction of drug effects, Boolean models and logic-based ODEs also enable the identification of key components in disease by using sensitivity analysis or \textit{in silico} knockdown experiments~\cite{Ryall2012,Khalilimeybodi2023_cardiomyopathy,Zeigler2020,RA_2021,Cancer_2}.
For example, the Ras GTPase signaling pathway has been found to show the greatest effect on myocyte size~\cite{Ryall2012}, where the increases of myocyte size are widely observed in cardiac hypertrophy.
Furthermore, proliferation and apoptosis networks have been identified to be associated with the survival rate of cancer patients~\cite{Cancer_2}.
Lastly, FOXO3 downregulation has been discovered as a potential mechanism of the alpelisib drug resistance for the estrogen receptor-positive (ER$^+$) PIK3CA-mutant breast cancer~\cite{GomezTe2021_Albert,GomezTe2017_Albert}.
In summary, choosing between network-based models and classical mechanistic models largely depends on the scientific question and the available data. 
Classic mechanistic models are most useful to understand detailed biochemical or biophysical processes underlying a system's behavior.
These models, based on known molecular mechanisms and interactions, provide deep insights into the dynamics and regulatory mechanisms of specific pathways, making them valuable for studying hypotheses on drug action. 
On the other hand, network-based models are preferred when studying large-scale interactions and relationships within biological systems, such as mapping out complex networks of genes, proteins, or signaling and metabolic pathways without delving into the detailed biochemical mechanisms. 
Furthermore, network-based approaches, which focus on connectivity and interaction patterns, excel in identifying drug targets (i.e., key species and interactions within a system), understanding system-wide behavior, and generating hypotheses about the roles of different components of this behavior. 
However, regardless of the chosen modeling method, constraining the model with experimental data to ensure it accurately reflects the biological system remains a gold standard, which increases confidence in the reliability of the model predictions.

\section{REDUCING UNCERTAINTY IN FREE PARAMETERS}
As noted above, a major challenge in model construction for systems biology and systems medicine is estimating the unknown model parameters~\cite{vittadelloOpenProblemsMathematical2022}.
In the previous section, we discussed how focusing on network topology rather than exact reaction kinetics can enable a network-level understanding of the system of interest, without predicting exact dynamics.
However, many applications require quantitative predictions of the dynamical response, and thus, accurate parameter estimates are required to constrain model predictions~\cite{Para_vitro_4}.
Many sources of uncertainty confound parameter estimation, including uncertainty in the model structure, the distribution of the model parameters, and the quality of the data used for model calibration~\cite{gerisUncertaintyBiologyComputational2016, Linden2022}.
However, the key challenge comes from the mismatch between the large number of free parameters and the relatively small number of observable species, which is known to lead to difficulties in successfully carrying out parameter estimation~\cite{gutenkunstUniversallySloppyParameter2007, monsalve-bravoAnalysisSloppinessModel2022}.
To enable parameter estimation, identifiability, and sensitivity analysis have become essential components of the systems biology toolkit to understanding the effects of parameters on model predictions and determining which parameters are important to constrain model behavior~\cite{Linden2022, qianSensitivityAnalysisMethods2020, mitraParameterEstimationUncertainty2019, wielandStructuralPracticalIdentifiability2021, monsalve-bravoAnalysisSloppinessModel2022, raueStructuralPracticalIdentifiability2009, ergulerPracticalLimitsReverse2011, gutenkunstUniversallySloppyParameter2007, sherQuantitativeSystemsPharmacology2022}.
Specifically, identifiability analysis determines whether parameters can uniquely be identified from available observations~\cite{wielandStructuralPracticalIdentifiability2021, anstett-collinPrioriIdentifiabilityOverview2020, ljungGlobalIdentifiabilityArbitrary1994}.
Sensitivity analysis, on the other hand, determines the contributions of model parameters to variability in the model predictions~\cite{qianSensitivityAnalysisMethods2020, saltelliGlobalSensitivityAnalysis2007, saltelliSensitivityAnalysisPractices2006}.
Here, we briefly discuss the state-of-the-art methods available for these two analyses and discuss how they are essential to effectively driving biological discovery from large models.

Identifiability analysis aims to find the subset of model parameters that can be uniquely estimated from the available data~\cite{wielandStructuralPracticalIdentifiability2021, hongGlobalIdentifiabilityDifferential2020, anstett-collinPrioriIdentifiabilityOverview2020, raueComparisonApproachesParameter2014, ljungGlobalIdentifiabilityArbitrary1994}.
Typically, identifiability can be broken into two components: i) structural identifiability analysis, which is performed before fitting the model to data~\cite{hongGlobalIdentifiabilityDifferential2020} and ii) practical identifiability analysis, which considers the quality of the data-fit~\cite{raueStructuralPracticalIdentifiability2009}.
\textit{A priori} structural identifiability analysis tests whether there is a unique map between parameters and the species that are observed in the available data~\cite{ljungGlobalIdentifiabilityArbitrary1994, hongGlobalIdentifiabilityDifferential2020}.
Methods for structural identifiability analysis rely on a range of mathematical theories, including observability (STRIKE-GOLD)~\cite{villaverde2016structural, villaverdeInputDependentStructuralIdentifiability2019}, differential algebra-based methods (DAISY software), generating series (GenSSI)~\cite{massonisFindingBreakingLie2020, chics2011genssi}, and randomized numerical algebra (SIAN and StructuralIdentifiability.jl)~\cite{hongSIANSoftwareStructural2019, hongGlobalIdentifiabilityDifferential2020}.
In general, these methods identify the parameters that can be uniquely identified from the set of observed model outputs.
Alternatively, \textit{a posteriori} practical identifiability considers the quality of the fit to data and defines identifiable parameters as those with well-constrained parameter estimates.
Methods for practical identifiability include the profile likelihood approach~\cite{raueStructuralPracticalIdentifiability2009}, and direct examination of Bayesian posterior densities~\cite{raueJoiningForcesBayesian2013, wielandStructuralPracticalIdentifiability2021}
While these approaches range from frequentist to Bayesian, they all similarly aim to analyze the width of the marginal predictive densities to infer the ability to estimate parameters with a high degree of certainty given available data.
We refer the reader to the following recent reviews for a more detailed overview of the theory and available methodology for identifiability analysis~\cite{wielandStructuralPracticalIdentifiability2021, anstett-collinPrioriIdentifiabilityOverview2020, raueComparisonApproachesParameter2014, barreiroBenchmarkingToolsPriori2022}.
Recent examples of \textit{a priori} structural identifiability analysis enabling accurate parameter estimation include the determination of identifiable subsets in a minimal physiologically-based pharmacokinetic model of the brain~\cite{dadashovaLocalIdentifiabilityAnalysis2024} and of commonly-included model \textit{motifs} are identifiable and suitable for building identifiable models~\cite{hausStructuralIdentifiabilityBiomolecular2023}
Furthermore, \textit{a posteriori} practical identifiable analysis provided important validation of parameter estimates in models of JAK2/STAT5 signaling~\cite{bachmannDivisionLaborDual2011}, Erythropoietin receptor signaling~\cite{beckerCoveringBroadDynamic2010}, and tumor growth~\cite{browningPredictingRadiotherapyPatient2024}.

Sensitivity analysis quantifies how variability in model parameters contributes to variability in model predictions~\cite{saltelliGlobalSensitivityAnalysis2007}.
The sensitivity of model outputs to variations in model parameters can be considered locally at a specific point in parameter space or globally across the entire space of plausible parameter values.
Local sensitivity analysis utilizes derivatives of model predictions with respect to model parameters evaluated locally at values of interest.
While local sensitivity can yield meaningful insights about optimal parameter estimates these results are often of limited value for the highly nonlinear models that we often encounter in systems biology and systems medicine~\cite{qianSensitivityAnalysisMethods2020, kentWhatCanWe2013}.
Alternatively, global sensitivity analysis decomposes the variance of model predictions into the contributions from each model parameter by varying those parameters over species ranges or distributions~\cite{qianSensitivityAnalysisMethods2020, saltelliGlobalSensitivityAnalysis2007}.
While a complete review of the methods for global sensitivity analysis is beyond the scope of this work, we note that Sobol's method~\cite{sobolGlobalSensitivityIndices2001}, Moris's method~\cite{morrisFactorialSamplingPlans1991}, and the Pearson Correlation Coefficient method~\cite{wentworthParameterSelectionVerification2016} have been successfully applied in systems biology.
The value of performing sensitivity analysis prior to parameter estimation is that sensitivity analysis can help to identify which parameters are most important to estimate to accurately constrain a model's predictions~\cite{Linden2022, wentworthParameterSelectionVerification2016}.
It has been shown that systems biology models often have many parameters that have little influence on model outputs---sloppy parameters---and a handful that strongly influence predictions---stiff parameters~\cite{gutenkunstUniversallySloppyParameter2007, monsalve-bravoAnalysisSloppinessModel2022}.
Therefore, focusing on estimating the stiff parameters can greatly improve the quality of the data fit and can reduce predictive uncertainty~\cite{Linden2022, monsalve-bravoAnalysisSloppinessModel2022, gutenkunstUniversallySloppyParameter2007}.

\begin{figure}
    \centering
    \includegraphics[width=\textwidth]{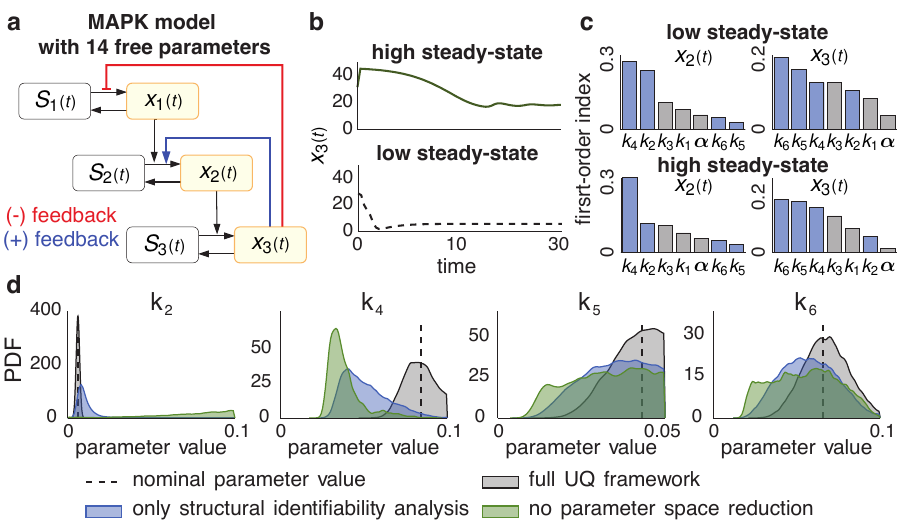}
    \caption{\textbf{Identifiability and sensitivity analyses enable parameter estimation of models from limited data.}
    (\textbf{a}) Phenomenological MAPK signaling model with 14 unknown model parameters.
    Model originally developed in~\cite{nguyenDYVIPACIntegratedAnalysis2015}.
    \textit{A priori} structural identifiability analysis showed that seven parameters are globally structurally identifiable~\cite{Linden2022} (results not shown here).
    (\textbf{b}) Example of bistable dynamics in $x_3(t)$ with a pre-defined set of nominal parameters.
    The low and high steady-states are reached by varying the initial conditions of the model.
    (\textbf{c}) Global sensitivity analysis shows that the steady-state concentrations of $x_2(t)$ and $x_3(t)$ are most sensitive to $k_2$, $k_4$, $k_5$, and $k_6$ (sensitive parameters shown in blue).
    (\textbf{d}) Reduction of the parameter space with both identifiability and sensitivity analyses is necessary to estimate unknown model parameters with a high degree of accuracy and certainty.
    Parameters are estimated using Bayesian inference from synthetic data generated from a simulation of the high steady-state.
    Green, estimated parameter probability densities for the top four parameters without any subspace reduction.
    Blue, estimated densities for the parameter space reduced by only applying structurally identifiability analysis.
    Black, estimates with the fully reduced parameter space using both identifiability and global sensitivity analysis.
    The certainty and accuracy of the estimated densities grow as the dimensionality of the parameter space is reduced.
    Panels a and b are adapted from Figure 4, and Panel d is adapted from Supplemental Figure 1 of Reference~\cite{Linden2022}.}
    \label{fig:SA-fig}
\end{figure}

To demonstrate how identifiability and sensitivity analyses can improve parameter estimation, we estimated the parameters of a small ODE-based model with and without these analyses~\cite{Linden2022}.
The model, a phenomenological representation of the Mitogen-activated protein kinase signaling pathway, has three state variables and 14 unknown parameters (Figure~\ref{fig:SA-fig}a)~\cite{nguyenDYVIPACIntegratedAnalysis2015}.
Under different combinations of parameters and initial conditions, the model can predict many dynamics, including bistability in $x_2(t)$ and $x_3(t)$ (Figure~\ref{fig:SA-fig}b).
First, the authors performed a global structural identifiability analysis using the SIAN toolbox~\cite{hongSIANSoftwareStructural2019} and found that seven parameters, $k_1$, $k_2$, $k_3$, $k_4$, $k_5$, $k_6$, and $\alpha$ were globally identifiable from measurements of all three states.
Next, they performed a global sensitivity analysis of these identifiable parameters using Sobol's method~\cite{saltelliGlobalSensitivityAnalysis2007} and found that four parameters $k_2$, $k_4$, $k_5$, and $k_6$, strongly influence $x_2(t)$ and $x_3(t)$ predicted steady-states (Figure~\ref{fig:SA-fig}c).
The authors then perform three separate rounds of parameter estimation using synthetic data generated by adding Gaussian noise to samples from the three states in the low steady-state in the bistable regime.
The nominal parameter values used to generate this data are indicated by dashed black lines in Figure~\ref{fig:SA-fig}d.
First, Linden et al. estimated all free model parameters, nine in total after excluding total concentration and integer-valued parameters.
Here, without any \textit{a priori} analysis, the estimated probability densities (green densities in Figure~\ref{fig:SA-fig}d) appear to be very wide and do not concentrate around the true nominal parameters.
Next, the authors estimated the reduced set of seven identifiable parameters.
Here, the estimated densities (blue densities in Figure~\ref{fig:SA-fig}d) begin to concentrate better around the nominal values; however, there is still a significant error in the estimate for $k_4$.
Lastly, the authors estimated the set of four identifiable and sensitive parameters.
Here, the estimated densities are all centered around the true nominal values and generally show significantly lower uncertainty than previous estimates.
In this example, reducing the set of free parameters to the identifiable set and then to the sensitive and influential set led to large improvements in the quality of parameter estimates and, thus, predictions.
Recent work by Linden-Santangeli et al.~\cite{linden-santangeliIncreasingCertaintySystems2024} has applied a similar framework of using identifiability and sensitivity analysis to enable parameter estimation for larger models with 10s-100s of free parameters.

As introduced here, \textit{a priori} identifiability and sensitivity analysis aim to reduce uncertainty in parameter estimates and, thus, increase certainty in model predictions.
These analyses are a part of the uncertainty quantification toolkit which aims to determine and account for uncertainties in modeling~\cite{smith2013uncertainty}.
Developed in the broader computational science community, rigorous uncertainty quantification beyond parameter estimation is beginning to become the standard practice in systems biology studies~\cite{gerisUncertaintyBiologyComputational2016, vittadelloOpenProblemsMathematical2022, sherQuantitativeSystemsPharmacology2022, miramsFickleHeartUncertainty2020}.
Recent work has shown that careful accounting of data and parametric uncertainty can improve model predictions and bring new insights to systems biology~\cite{Linden2022,simpsonParameterIdentifiabilityParameter2024, huberSystematicBayesianPosterior2023, caoQuantificationModelData2020b, wangQuantificationUncertaintyNew2020, irvinModelCertaintyCellular2023}.
Furthermore, methods to account for model uncertainty and select models from a set of candidates can further improve model-based predictions~\cite{linden-santangeliIncreasingCertaintySystems2024, liuParameterIdentifiabilityModel2023, shockleySignalIntegrationInformation2019, kirkModelSelectionSystems2013}.
Lastly, global sensitivity analysis can reveal important components of and interactions within a system without explicitly fitting the model to data~\cite{saltelliGlobalSensitivityAnalysis2007}. 
As models become a more routine component of studying biological and physiological systems, end-to-end uncertainty quantification that accounts for all sources of uncertainty is essential to improving confidence in and enabling rigorous statistical analyses of predicted outcomes.

\section{INTEGRATED SIGNALING-GENE-METABOLIC NETWORKS AND VALIDATIONS WITH OMICS DATA}
While signaling, gene regulatory, and metabolic networks are interconnected and work together to execute cell functions, integrating two or all of these three layers of networks \textit{in silico} has attracted increasing attention.
Integrated models allow studies on the interactions between the different networks and thus can predict new disease mechanisms and novel therapeutic strategies~\cite{Pollizzi2014_review_meta_sign,Imam2015_review_gene_meta,Osterberg2021_SMG,Rodriguez2016_SM}.
Here, we focus on the integration of all three types of networks~\cite{Tenazinha2011_review_SMG,Burke2020_SMG} and review several recent applications.
One advantage is that not only are more molecules predicted compared with those for only one type of network, but the prediction accuracy also increases.  
For example, Wu et al. integrated signaling, gene regulatory, and metabolic networks in the liver and then predicted the effect of cortisol infusion on the glucose, lactate, and Cytochrome P450 3A4 (an enzyme that is responsible for the metabolic clearance)~\cite{Wu2016_MUFINS};
Furthermore, Covert et al.~\cite{Covert2008_SMG} integrated metabolic, transcriptional regulatory, and signal
transduction models in \textit{Escherichia coli} and obtained a higher prediction accuracy for metabolites and transporters compared with the model only integrating metabolic networks and transcriptional regulation.

To provide an outline of how to construct an integrated model of signaling, gene regulation, and metabolism, we focus on one recent study from our group~\cite{Khalilimeybodi2024}.
In this study, we developed a computational model that integrates signaling, metabolic and gene regulatory networks for hypertrophic cardiomyopathy (\textbf{Figure \ref{fig:whole_cell}a}).
The signaling and gene regulatory networks were modeled by the stochastic version of a logic-based differential equation, where the dynamics of each species are governed by 
$dy/dt=\frac{1}{\tau_y}\Big[F(x,y,z,...)y_{max}-y+\eta\Big]$.
Here, $\eta$ is a process noise term with a mean of 0 that captures the biological noise in cells, including stochasticity of chemical reactions and varied cellular environments. 
The standard deviation of $\eta$ is usually determined empirically.
The metabolic network was modeled using \textit{iCardio} and was originally developed by Dougherty et al.~\cite{Dougherty2021}. 
The coupling between signaling, metabolic, and gene regulatory networks is achieved by introducing the following regulations: i) the components in signaling network regulate the level of transcription factors (TFs; yellow circles in \textbf{Figure \ref{fig:whole_cell}a}); ii) mRNA (red rectangles in \textbf{Figure \ref{fig:whole_cell}a}) in the gene regulatory network is translated into proteins and thus increases the level of proteins in the signaling pathway; iii) the mRNA levels in the gene regulatory network are inputs to the iCardio model by assuming a linear association between mRNA expression and protein levels (yellow arrow labeled by ``Metabolic Enzymes Regulation'' in \textbf{Figure \ref{fig:whole_cell}a}); iv) the components in signaling network enzymatically regulate the metabolites, e.g., post-translational regulation in \textbf{Figure \ref{fig:whole_cell}a}; v) metabolites such as adenosine triphosphate (ATP) are involved in signaling pathways and react with other signaling components (blue arrow labeled with `` Metabolic Changes" in \textbf{Figure \ref{fig:whole_cell}a}).
The above couplings help build an integrated model that considers almost all interactions between different layers of networks involved in hypertrophic cardiomyopathy.

Validation of the integrated model is achieved by comparing the signaling activities and gene expressions at steady state to existing experimental data of the transcriptomes and signaling molecules (\textbf{Figure \ref{fig:whole_cell}b}).
First, two different sets of kinetic parameters are determined: one set of default parameters that mimic the healthy control setting and the other set that corresponds to the disease state.
Then, multiple \textit{in silico} replicates are simulated for the integrated model with both sets of kinetic parameters. 
Next, the steady-state values of signaling molecules and mRNAs are recorded (bottom left form in \textbf{Figure \ref{fig:whole_cell}b}), where the mRNA data can be regarded as \textit{in silico} transcriptomes.
For each species, an unpaired t-test between healthy control and disease state is performed, and the corresponding p-value is employed to compute the false discovery rate (FDR) (also called FDR adjusted p-value)~\cite{Benjamini1995}.
The fold change is the ratio of the mean difference of species activities between the healthy control and disease state to the mean species activity under the healthy control condition. 
Then, the FDR and fold change are used to determine the trend of species activities: if FDR is larger than the threshold (usually set to be 0.05), the species is assumed to be no change between healthy control and disease state; if FDR is smaller than the threshold, the species is assumed to be ``increase'' (or ``decrease'') in the disease state when the $log_2$(fold change) is larger (or smaller) than 0.
Thus, the \textit{in silico} list of differentially expressed signaling molecules and genes is generated (top right form in \textbf{Figure \ref{fig:whole_cell}b}).
To compare this list with the experimental data, the experimental data are also sorted into three groups based on their statistical significance against controls:  decrease, no change, or increase.
One example of these three groups is the differentially expressed genes from transcriptome data.
If most of the species show the same trend between experimental data and model prediction, then the model is considered to be experimentally validated.

One primary advantage of the integrated hypertrophic cardiomyopathy model is that although no parameter estimation is required, the model is still able to fit the static qualitative trends observed in the experimental data.
Eliminating the need for direct parameter estimation also reduced the computational demands of developing such a large model.
Nevertheless, the model was only able to capture the time-independent qualitative trends in the data and was unable to make time-dependent or quantitative predictions. 
Another advantage of developing integrated models is their ability to identify drug targets in complex diseases where cell response significantly depends on its context. 
In hypertrophic cardiomyopathy, cell context includes experimental settings (e.g., in vitro vs. in vivo), environmental factors (e.g., comorbidity), gene mutations, cellular noise, extracellular matrix (ECM) structure (e.g., stiffness), and cell stimuli (type and frequency), with variations in these factors affect disease progression~\cite{Maron2019-kx}. 
The integrated model provides a computational framework to account for many of these contexts in the modeled system~\cite{Ji2017}.

\begin{figure}[H]
  \centering
  \includegraphics[width=1\textwidth]{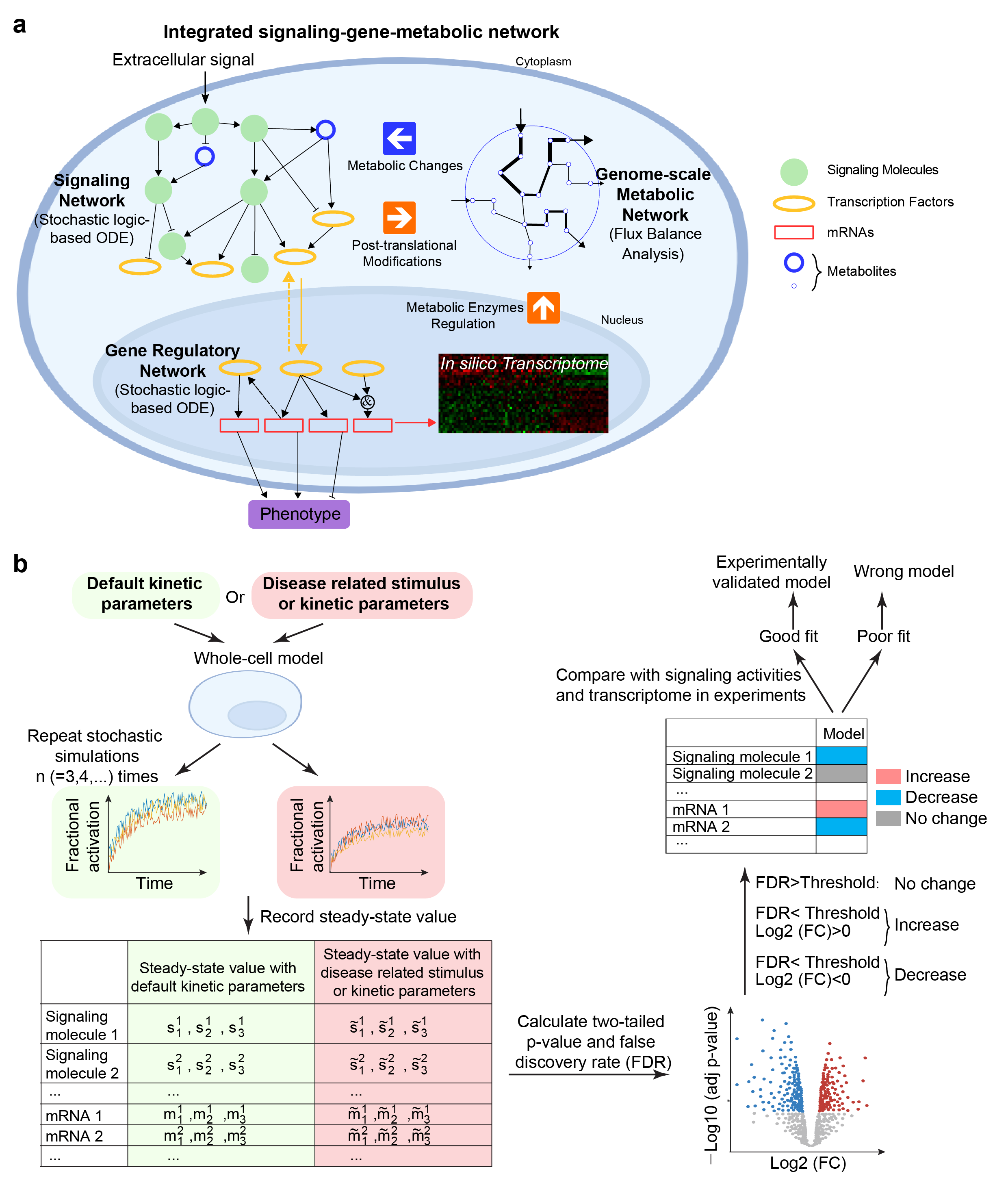}
\end{figure}
\begin{figure}[H]
\captionof{figure}{\textbf{Schematic of integrated signaling-gene-metabolic networks and model validations with omics data.} 
(\textbf{a}) Schematic of the integrated signaling-gene-metabolic network. 
The signaling and gene regulatory networks are modeled by stochastic logic-based ODEs, and the metabolic network is modeled by flux balance analysis. 
These three types of networks are coupled by transcription factors regulated by the signaling network, protein production caused by gene expression, metabolic enzyme regulation, post-translational modifications, and chemical reaction alterations caused by metabolic changes. 
(\textbf{b}) Schematic of model validations with omics data. 
Two distinct sets of kinetic parameters are used to mimic the healthy and disease conditions. 
For each set of parameters, the model is simulated multiple times to compute the steady-state values and their statistics.
The trend of steady-state values is compared to experimental data on the changes in signaling activities and mRNA levels. 
The good fit to data suggests a model that is well-validated by experiment.
The schematic of cell and volcano plot were generated by \textit{BioRender}.
Panel a is adapted from Figure 1 of Reference~\cite{Khalilimeybodi2024}.
}
  \label{fig:whole_cell}
  \centering
\end{figure}

\section{INTEGRATED MODELS GENERATE INSIGHTS INTO NETWORK REDUCTION AND DRUG EFFICACY}
After validating the integrated model with omics data, such model is a powerful tool for providing insights into disease and predicting the efficacy of drug treatment. 
One key insight is the identification of the key reactions in disease, which is achieved by sensitivity analysis~\cite{Khalilimeybodi2024,Fowler2024}. 
In this analysis, the strength of each regulatory link is perturbed, and then the corresponding model output is calculated.
The quantitative sensitivity metric (e.g., the Morris sensitivity index) reflects the impact of perturbations on the model output and thus can be used to rank the reactions.
This approach can help to reduce the complexity of the original cellular network to the core regulations that contribute to the diseased state. 
Another important insight of the integrated model is to generate \textit{in silico} predictions of drug efficacy. 
In general, if the drug inhibits (or activates) the activity of target molecules, the level of the target molecules in the integrated model is set to decrease (or increase in the case of activation) to mimic the effect of the drug of interest. 
Then the model is simulated given these changes, and outputs such as the cellular phenotype or the level of specific molecules can be compared with that in the absence of the introduced \textit{drug}.
In this way, the effect of drugs can be predicted, and drug combinations of drugs can be explored to improve potential therapeutic options.
In this section, we introduce three examples of network models that have identified key intracellular reactions and predicted potential combinations of drug targets.

\subsection{Applications in Hypertrophic Cardiomyopathy (HCM)}
To date, many computational studies, including multiscale models, have investigated the mechanisms of hypertrophic cardiomyopathy (HCM) at molecular, cellular, and organ levels~\cite{Margara2022-nz,Campbell2011-ze,Aboelkassem2019-pb}. 
Most of these studies have focused on the question of how HCM mutation in sarcomere genes affects cardiac contractility and contributes to arrhythmogenesis~\cite{Doh2019-mc,Vera2019-bo,Mijailovich2017-pt,Zile2017-kb} rather than cardiac growth and remodeling in HCM.
However, in a recent study, Davis et al.~\cite{Davis2016-ra} introduced a new model for myocardial growth based on a 'tension index' determined from cardiac twitch computational models, and showed that changes in the tension-time integral correlate with the type and severity of myocardial remodeling in HCM and dilated cardiomyopathy (DCM) hearts. 
They also found that while calcineurin-NFAT signaling regulates the extent of cardiac hypertrophy, MEK-ERK1/2 signaling determines the growth direction by promoting the addition of sarcomeres in parallel for cardiomyocyte thickening, whereas inhibiting MEK-ERK1/2 leads to cardiomyocyte elongation by adding sarcomeres in series~\cite{Davis2016-ra}. 
Based on this study, we developed an integrated model to predict cardiomyocyte responses to HCM mutations across various signaling, transcriptional, and metabolic levels~\cite{Khalilimeybodi2024}.

By employing a global sensitivity analysis, we identified the key reactions that affect the gene expression levels in hypertrophic cardiomyopathy (\textbf{Figure \ref{fig:insights}a}).
Given the differences between in vivo mouse models of hypertrophic cardiomyopathy and HCM patients, three types of transcriptomic data were simulated (three bars in \textbf{Figure \ref{fig:insights}a}): non-obstructive R403Q-\textalpha MyHC in mouse, R92W-TnT HCM mutations in mouse, and human transcriptomic datasets (GSE36961: mRNA, GSE36946: miRNA).
We revealed that cardiomyocyte response in hypertrophic cardiomyopathy is directed by a mix of shared and context-specific reactions. 
The shared reactions across the three contexts included AMPK activating PGC1\textalpha, titin activating FHL1, and AMPK regulating ATP/ADP levels, with AMPK itself being activated by LKB1 and inhibited by PI3K/AKT.
We also identified some reactions specific to each context: for the \textalpha MyHC mutation, interactions such as ROS production by NOX4, PKD activation by PKC, the activation of PPAR by mTOR, and the regulatory reactions linking \ce{Ca^2+} transients to the sarcomere active force were central in controlling the cardiomyocyte response; for the TnT mutation, major regulatory reactions included regulation of \ce{Ca^2+} diastolic level by PLB through SERCA, NF\textkappa B regulation by PPAR and PI3K/AKT, CaMK activation by ROS, and activation of Ras through growth factor receptors; for HCM patients, major reactions were a combination of those in \textalpha MyhC and TnT contexts.

Moreover, we screened potential drug targets in HCM by performing a combinatory perturbation analysis using the integrated model~\cite{Khalilimeybodi2024}.
The effects of six therapeutic strategies for HCM cardiomyocytes were predicted (\textbf{Figure \ref{fig:insights}b}): a \ce{Ca^2+} sensitivity reduction only, the \ce{Ca^2+} sensitivity reduction paired with ATP level decrease, ROS inhibition, TF53 inhibition, AMPK inhibition, or AMPK hyperactivation. 
Treatment with only \ce{Ca^2+} sensitivity reduction resulted in a significant decrease in the hypertrophic growth index, but no notable change in the apoptosis index compared to the untreated case (the first and second bars in \textbf{Figure \ref{fig:insights}b}).
We examined combination treatments with \ce{Ca^2+} sensitivity reduction (\textbf{Figure \ref{fig:insights}b}), finding ATP level reduction had no significant impact, ROS inhibition reduced both indexes, TP53 inhibitor decreased apoptosis but increased hypertrophic growth, AMPK inhibition raised both indexes and AMPK hyperactivation showed no significant effect.

Of all the drug combinations, \ce{Ca^2+} sensitivity reduction paired with ROS inhibition is most effective in reducing both hypertrophic growth and apoptosis indexes. Consequently, the effects of this combination, along with \ce{Ca^2+} sensitivity reduction alone and paired with AMPK hyperactivation, were predicted for cardiomyocyte metabolism (\textbf{Figure \ref{fig:insights}c}). 
The HCM mutation was predicted to decrease fatty acid metabolism and increase the creatine kinase system and carbohydrate metabolism (data not shown here; see~\cite{Khalilimeybodi2024} for more details). The three treatments were predicted to reverse these effects, upregulating fatty acid metabolism and downregulating the creatine kinase system and carbohydrate metabolism (\textbf{Figure \ref{fig:insights}c}). The combination of \ce{Ca^2+} sensitivity reduction and ROS inhibition proved to be more effective than the other two, offering promising targets for future drug development for HCM.

This study emphasizes the advantages of network modeling in developing integrated models for drug discovery. Such models offer a comprehensive understanding of cellular interactions, identify potential drug targets more effectively, and accelerate the discovery and development process by simulating drug effects on complex biological systems.

\subsection{Applications in Exercise}
Using the network modeling approach, Fowler et al.~\cite{Fowler2024} developed an integrated model to predict the differential phenotypic responses of skeletal myocytes to resistance and endurance exercise. 
This model was used to predict changes in 12 phenotypic outcomes in response to exercise inputs and accurately forecast 85\% of resistance and 75\% of endurance exercise measurements from independent studies. 
All phenotypic outputs responded to both exercise types, but with varying magnitudes; the model specifically predicted differences in gene activity related to inflammation, protein synthesis, cell growth, and protein degradation between resistance and endurance exercise~\cite{Fowler2024}.
Sensitivity analysis highlighted key pathways that regulate responses to both exercise forms, including MAP kinase, PI3 kinase, STARS, NF{\textkappa}B, cyclic AMP, and calcium.
More specifically, the analyses predicted that resistance exercise mainly activates cell growth and protein synthesis via mTOR signaling, while endurance exercise activates inflammation through NF{\textkappa}B and ROS. 
Inhibiting TNF{\textalpha} reduced differences in protein synthesis between exercise types. The model also revealed that inhibiting ROS affects protein synthesis during endurance but not resistance exercise. 
However, the model could not predict the expected preferential activation of mitochondrial biogenesis by endurance exercise, as PGC1{\textalpha} activation was counteracted by NF{\textkappa}B and PKC activities. 
By simulating multiple training scenarios, they showed that protein synthesis, cell growth, and anti-inflammatory activity increased more with concurrent training than with endurance training alone but less than with resistance training alone. 
Simulating ROS knockdown reduced the effects of endurance and concurrent training on protein synthesis while slightly increasing the effects of resistance training. Knocking down TNF{\textalpha} reduced the effect of resistance training on protein synthesis to levels similar to endurance training. They suggested that TNF{\textalpha} activation of MAPK signaling, S6K, and rpS6 is crucial for regulating protein synthesis responses to different exercise types, while AMPK knockdown had minimal impact on these differences~\cite{Fowler2024}.
This integrated model of skeletal muscle cells can benefit future interventions by predicting differential responses to various exercise prescriptions and optimizing personalized training regimens for specific phenotypic outcomes.

\begin{figure}[H]
  \centering
  \includegraphics[width=1\textwidth]{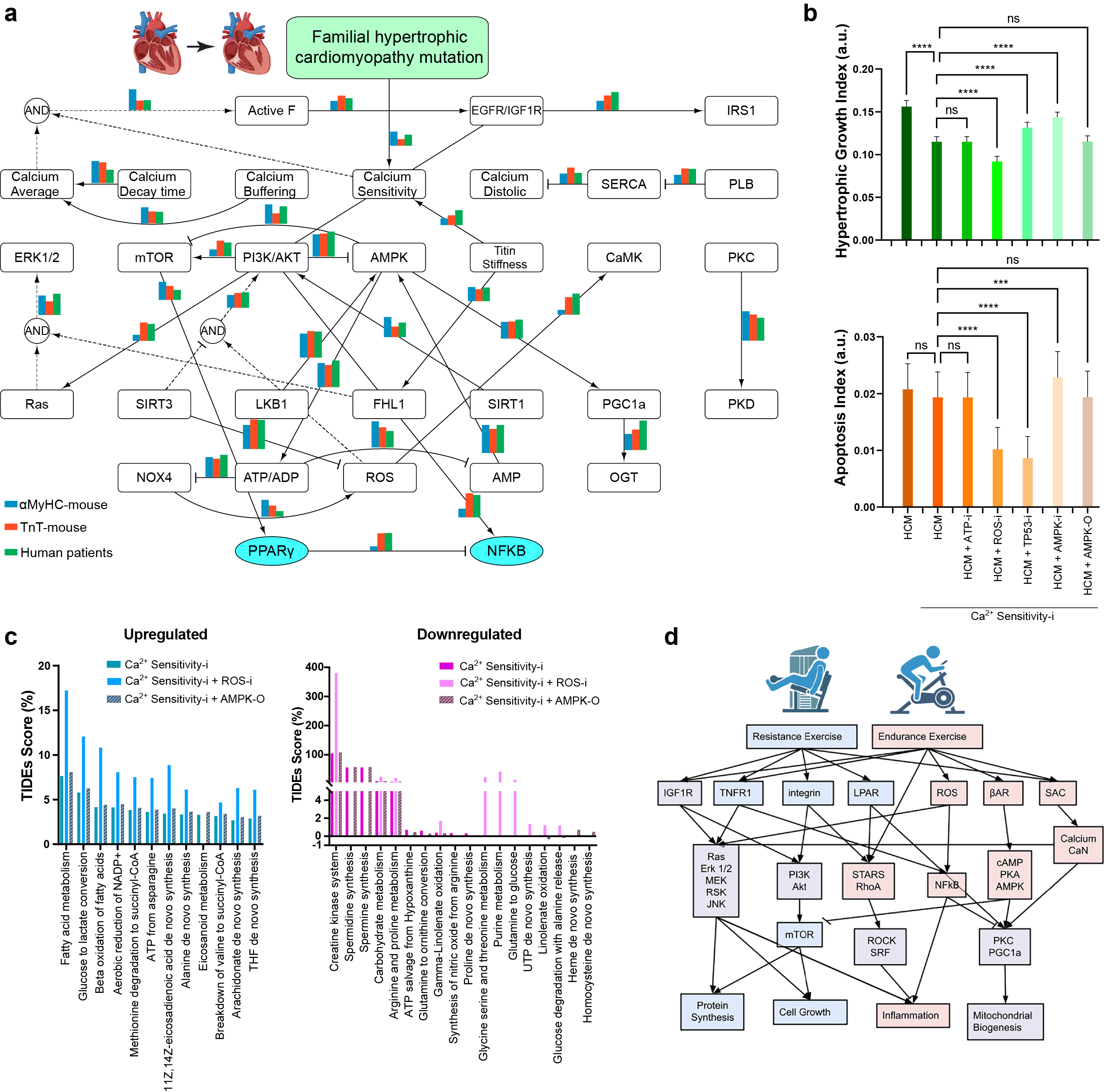}
\end{figure}
\begin{figure}[H]
\captionof{figure}{\textbf{Prediction of major regulatory reactions and potential drug targets in familial hypertrophic cardiomyopathy (HCM) and exercise.} (\textbf{a}) Context-dependent regulatory reactions controlling cardiomyocyte response in three HCM contexts: mouse R403Q-MyHC, mouse R92W-TnT, and human HCM patients. In-silico prediction of the impact of model-informed drug targets on (\textbf{b}) cardiomyocyte hypertrophic growth and apoptosis and (\textbf{c}) metabolic functions. (\textbf{d}) Model-informed key nodes and pathways regulating skeletal muscle response to endurance and resistance exercise.
The schematic of heart and exercise were generated by \textit{BioRender}.
Panel a is adapted from Figure 6, and Panels b-c are adapted from Figure 7 of Reference~\cite{Khalilimeybodi2024}. 
Panel d is adapted from Figure 6 of Reference~\cite{Fowler2024}.
}
  \label{fig:insights}
  \centering
\end{figure}

\subsection{Applications in Cell Survival and Cancer}
Integrated models can also simulate cell proliferation, making them useful for investigating cell survival in cancers.
One hallmark of cancer is sustained and aberrant cellular proliferation that is induced by secreted growth factors. 
Scientists have explored how the homeostasis of cell populations is achieved by building tissue-scale models, which consist of cell numbers and growth factor concentrations~\cite{Alon2018_homeostasis,Alon2018_cell,Hart2014_Homeostasis}. 
Since the growth factors regulate many intracellular pathways, the coupling of these intracellular pathways to the previously developed tissue-scale models enables the understanding of the relationship between tissue homeostasis and secreted growth factors in both cellular and tissue scales. 
The classical intracellular pathways activated by growth factors include mitogen-activated protein kinase (MAPK) cascade, phosphatidylinositol polyphosphate (PIP) signaling, and so on~\cite{Oda_EGFR_review,Schoeberl2002_EGFR_model}. 
Moreover, recent studies revealed that growth factors also activate a circuit at the Golgi which is composed of two types of GTPase (monomeric GTPase Arf1 and heterotrimeric GTPases Gi), and this circuit is able to regulate the role of Golgi in the autocrine secretion of growth factors~\cite{Lo2015_GIV,Qiao2023_MSB,Qiao2023_npj}.
Thus, the circuit of coupled GTPases at the Golgi closes the loop between growth factor sensing and secretion.
By using logic-based ODEs to model this circuit and then coupling this model to the tissue homeostasis modeling~\cite{Qiao2023_MSB}, we not only reproduced the experimentally observed dynamics of signaling molecules but also predicted the role of coupling Arf1 and Gi in the cell proliferation.
We anticipate that in the near future, the community will be able to build integrated models based on multi-omics data~\cite{Hasin2017_review_multi_omics,Gonzalez2022_Crohn} for complex families of diseases such as cancer, which are as detailed as those for hypertrophic cardiomyopathy.

\section{FUTURE OUTLOOK}
Here, we briefly summarize opportunities for computational modeling and uncertainty quantification to make continued impacts in systems biology and systems medicine. 
   \subsection{Integrating Systems Biology with Modern Machine and Statistical Learning Software}
   In this work, we outlined \textit{a priori} identifiability and sensitivity analysis to enable parameter estimation. 
   Alternatively, exploiting the geometry of loss or likelihood functions by using information about their gradients or curvatures to guide sampling to the most identifiable directions in parameter space---the directions with the greatest curavture~\cite{gutenkunstUniversallySloppyParameter2007, monsalve-bravoAnalysisSloppinessModel2022}---can improve the efficiency and certainty of parameter estimation.
   Outside of systems biology modern software for machine learning and statistical inference, such as PyTorch~\cite{paszkePyTorchImperativeStyle2019}, Jax~\cite{jax2018github}, and PyMC~\cite{pymc2023}, already make use of gradient- and Hessian-informed algorithms, such as stochastic gradient-descent and quasi-Newton algorithms~\cite{kutz2013data}, and Hamiltonian Markov chain Monte Carlo~\cite{brooksHandbookMarkovChain2011}.
   However, most systems biology software limits the usage of these algorithms, because they do not provide the ability to efficiently and accurately compute gradients of model predictions with respect to input parameters.
   Future development of new systems biology software that interfaces with modern backends for auto-differentiation and statistical learning has the potential to further improve how we calibrate models in systems biology.
   Early headway into these efforts could revolve around the utilization of the auto-differentiation features provided by the Julia programming language~\cite{roeschJuliaBiologists2021} and the Python Jax ecosystem~\cite{jax2018github, kidger2021on}.
   Future versions of systems biology software, such as VCell~\cite{Schaff1997_vcell1, COWAN2012195_Vcell2} or COPASI~\cite{COPASI}, would greatly expand the tools available to systems biologists and could automate detection and mitigation of parameter non-identifiability without requiring additional \textit{a priori} analyses.
   
   \subsection{Accounting for Model Uncertainty with Multimodel Inference}
    In almost all biological systems, many related mathematical models exist that vary in the simplifying assumptions used to represent the biological process mathematically.
    For example, the MAPK model presented in Figure~\ref{fig:SA-fig} from~\cite{nguyenDYVIPACIntegratedAnalysis2015} represents the signaling pathway with phenomenological equations, while additional models represent the pathway with varying levels of physiological detail~\cite{Orton2005}.
    One should account for the uncertainties associated with these assumptions when making predictions~\cite{smith2013uncertainty}.
    Model selection has been the preferred approach to select a single ``best'' model when multiple models are available~\cite{kirkModelSelectionSystems2013, Burnham2002-rl}.
    However, given the limited and noisy data in systems biology and medicine, these approaches may lead to biases and misrepresentations of uncertainty due to selecting a single model~\cite{Burnham2002-rl,Vehtari2016-sb}.
    Multimodel inference (MMI)~\cite{Burnham2002-rl,Vehtari2016-sb} leverages the entire set of available models to avoid selection biases and account for model form uncertainty.
    Recently, we applied Bayesian multimodel inference to a set of ten models of the MAPK signaling pathway and showed that considering all ten models together improves predictive certainty and can lead to new discoveries of the mechanistic underpinnings of observed signaling phenomena~\cite{linden-santangeliIncreasingCertaintySystems2024}.
    Future applications in systems biology and medicine should emphasize testing modeling assumptions and should apply MMI tools when multiple models of the same system are available. 
   
   \subsection{Integrating Models with Disease in QSP and PBPK Models}
   
   Integrated models, particularly those highlighted in Quantitative Systems Pharmacology (QSP) and Physiologically Based Pharmacokinetic (PBPK) models, have revolutionized our understanding and approach to disease modeling~\cite{aghamiri2022recent}. 
   QSP models are mechanistic in nature, aiming to simulate the complex interactions between biological systems and pharmacological agents by integrating detailed descriptions of molecular, cellular, and systemic processes~\cite{azer2021history}.
   QSP models help bridge the gap between preclinical findings and clinical outcomes by capturing the dynamics of drug action and disease progression across various biological scales~\cite{bradshaw2019applications}. 
   For instance, QSP models can simulate how a drug interacts with multiple targets within a biological pathway~\cite{bai2019translational}, predict therapeutic and adverse effects~\cite{kadra2018predicting}, and explain the variability in patient responses through virtual population models~\cite{cheng2017qsp, cheng2022virtual}.
   QSP models can capture the enhanced avidity, altered signaling pathways, and biological effects of bivalent antibodies' crosslinking and receptor clustering, predicting their potential advantages in potency and duration of action over monovalent antibodies~\cite{lesley1984selection}. 
   PBPK models complement QSP models by providing a framework to describe the absorption, distribution, metabolism, and excretion (ADME) processes of drugs within the body~\cite{shebley2018physiologically}.
   PBPK models allow for the prediction of drug concentration profiles in different tissues and organs, facilitating dose optimization and individualized therapy~\cite{kuepfer2016applied}. By incorporating detailed physiological, anatomical, and biochemical data, PBPK models can simulate drug kinetics in virtual populations, accounting for variability due to factors like age, gender, and genetic differences~\cite{huisinga2012modeling}. 
   The integration of QSP and PBPK models enables a comprehensive systems pharmacology approach, where the pharmacokinetics and pharmacodynamics of drugs are interconnected within a unified framework~\cite{joshi2023convergence}.

\section*{DISCLOSURE STATEMENT}
The authors are not aware of any affiliations, memberships, funding, or financial holdings that
might be perceived as affecting the objectivity of this review. 

\section*{ACKNOWLEDGMENTS}
AK acknowledges support from an American Heart Association Postdoctoral Fellowship (ID: 898850).
NJL acknowledges support from the National Institute of Biomedical Imaging and Bioengineering (NIBIB) of the National Institutes of Health (NIH; \url{https://www.nibib.nih.gov}) under award number T32EB9380 and a UCSD Sloan Scholar Fellowship from the Alfred P. Sloan Foundation (\url{https://sloan.org}). 
PR acknowledges support from Air Force Office of Scientific Research (AFOSR; \url{https://www.afrl.af.mil/AFOSR/}) Multidisciplinary University Research Initiative (MURI) grant FA9550-18-1-0051.
We acknowledge additional support from the Wu Tsai Human Performance Alliance and the Joe and Clara Tsai Foundation.


\end{document}